\documentclass[turnofflinenumbers]{JASA-EL}

\usepackage{tikz,pgfplots}
\usetikzlibrary{decorations.pathreplacing,plotmarks}
\pgfplotsset{compat=1.7}
\begin{document}
{\it The following article has been submitted to The Journal of the Acoustical Society of America Express Letters. After it is published, it will be found at http://asa.scitation.org/journal/jel.}

\bigskip

%% Version in square brackets will appear at top right of all pages except title page:
\title[Individualized sound pressure equalization]{Individualized sound pressure equalization in hearing devices exploiting an electro-acoustic model}

%% Use as many times as you want to:
\author{Henning Schepker}
\altaffiliation{Now at Starkey Hearing Technologies, Eden Prairie, MN 55344, United States}
\email{henning.schepker@uni-oldenburg.de}
\correspondingauthor
\affiliation{Department of Medical Physics and Acoustics and Cluster of Excellence "Hearing4all", University of Oldenburg, Germany}

\author{Reinhild Roden}
\email{reinhild.roden@jade-hs.de}
\affiliation{Institut f\"ur H\"ortechnik und Audiologie, Jade Hochschule, Oldenburg, Germany}

\author{Florian Denk}
\altaffiliation{Now at German Institute of Hearing Aids, L{\"u}beck, Germany.}
\email{florian.denk@uni-oldenburg.de}
\affiliation{Department of Medical Physics and Acoustics and Excellence cluster "Hearing4all", University of Oldenburg, Germany}

\author{Birger Kollmeier}
\email{birger.kollmeier@uni-oldenburg.de}
\affiliation{Department of Medical Physics and Acoustics and Excellence cluster "Hearing4all", University of Oldenburg, Germany}

\author{Matthias Blau}
\email{matthias.blau@jade-hs.de}
\affiliation{Institut f\"ur H\"ortechnik und Audiologie, Jade Hochschule, Oldenburg, Germany}

\author{Simon Doclo}
\email{simon.doclo@uni-oldenburg.de}
\affiliation{Department of Medical Physics and Acoustics and Excellence cluster "Hearing4all", University of Oldenburg, Germany}

%% For example:
% \author{Author One}
% \altaffiliation{Also at: Another University, City, State, Zipcode, Country.}
% \email{author.one@someplace.edu}
% \author{Author Two}
% \email{author.two@someplace.edu}
% \affiliation{Department1,  University1, City, State ZipCode,
% Country}
% \author{Author Three}
% \email{author.three@someplace.edu}
% \correspondingauthor
%% This command must be used for corresponding author. Enter it following
%% the author email to whom email should be sent:
%%

%% Optional
\date{\today} 

\begin{abstract}
To improve sound quality in hearing devices, the hearing device output should be appropriately equalized. To achieve optimal individualized equalization typically requires knowledge of all transfer functions between the source, the hearing device, and the individual eardrum. However, in practice the measurement of all of these transfer functions is not feasible. This study investigates sound pressure equalization using different transfer function estimates. Specifically, an electro-acoustic model is used to predict the sound pressure at the individual eardrum, and average estimates are used to predict the remaining transfer functions. Experimental results show that using these assumptions a practically feasible and close-to-optimal individualized sound pressure equalization can be achieved.
\end{abstract}

%% pacs numbers not used

\maketitle

%  End of title page for JASA-EL --------------------------------- %

\section{Introduction}
In recent years, major improvements have been made in the area of assistive hearing devices such as hearing aids, consumer headsets and hearables. However, the sound quality has been identified as one of the limiting factors for the end-user acceptance \cite{Killion2004,Sockalingam2009}. In order to improve the sound quality, sound pressure equalization algorithms can be used that aim at perceptually restoring the open-ear characteristics when a hearing device is used, typically referred to as acoustic transparency. To achieve acoustic transparency, different approaches can be used \cite{Hoffmann2013,Denk2018,Vaelemaeki2015,Gupta2019,Fabry2019,Schepker2020jasm}. Most of these approaches rely on the availability of all acoustic transfer functions between the source(s), the hearing device and the individual eardrum to compute the equalization filter. Thus, an understanding of the impact of different estimations of the required acoustic transfer functions is essential for the practical application of individualized sound pressure equalization. 

A simple way to circumvent the problem of measuring all required acoustric transfer functions for a user is to use a generic filter that is precomputed using, e.g., measurements on a dummy head. However, a precomputed equalization filter does not match the individual ear acoustics and can thus lead to a degraded perceived sound quality \cite{Schepker2020jaes}. Accurate estimates of the transfer function to the individual eardrum could be obtained using probe-tube measurements \cite{Hellstrom1993}. However, probe-tube measurements are time-consming, delicate and thus difficult to obtain, which makes them unsuitable for many practical applications. Assuming that the pressure inside the ear canal does not vary substantially allows to use an in-ear microphone in a hearing device, e.g., a microphone placed at the inner face of an earmould. It is well-known that the sound pressure in the ear canal can vary substantially between the position of the eardrum and different positions along the ear canal \cite{Stinson1985,McCreery2009}, and hence also the position the an in-ear microphone of a hearing device. Therefore, electro-acoustic models have been proposed to estimate the sound pressure at the individual eardrum based on an in-ear microphone \cite{Hudde1999,SankowskyRothe2015,Vogl2019}. In this paper we consider a custom in-the-ear multi-microphone earpeice, as presented in \cite{Denk2018}, and use an estimate of the transfer between the receiver of the hearing device and the individual eardrum obtained from an electro-acoustic model of the hearing device and the individual ear canal \cite{Vogl2019}. Furthermore, we propose to precompute the required transfer functions between the source and the occluded ear as well as the source and the open ear based on multiple measurements obtained from different subjects, similarly as done in \cite{Denk2018TIH} for the open ear. This effectively allows to individualize the equalization filter with minimal measurement effort.

\section{Acoustic Transparency}
\begin{figure}
 	\centering
 		\begin{scriptsize}
		\begin{tikzpicture}
		%%%% source loudspeaker %%%%%%
		\draw (7.1,-2.45) rectangle (7.2,-2.05);
		\draw (7.1,-2.45) -- (6.95,-2.6) -- (6.95,-1.9) -- (7.1,-2.05);
		% draw connections
		\draw[->] (6.7,-2.25) -- (5.0,-2.25);
		\draw[->] (6.7,-2.25) -| (6.0,-1.00);
		\node at (6.25,-2.45) {$s[k]$};
		%%%% eardrum "mic" %%%%%%
		\draw (6.7,1.25) circle (0.1);
		\draw (6.6,1.1) -- (6.6,1.4);
		\draw (6.8,1.25) -- (7.0,1.25);
		\node at (6.95,1.55) {eardrum};
		\node at (6.95,0.95) {$t_{aid}[k]$};
		%%%% loudspeaker %%%%%%	
		% first loudspeaker
		\draw (3,1.45) rectangle (3.1,1.05);
		\draw (3.1,1.45) -- (3.25,1.6) -- (3.25,0.9) -- (3.1,1.05);
		% loudspeaker beamformer
		\draw (1.9,0.75) rectangle (0.9,1.75);
		\node at (1.4,1.25) {$A(q)$};
		% loudspeaker beamformer to loudspeakers
		\draw[->] (1.9,1.25) -- (3.0,1.25);
		%%%% hearing device processing %%%%%%	
		% gain rectangle
		\draw (-0.5,0) rectangle (0.5,-1.0);
		\node at (0,-0.5) {$G(q)$};
		\node at (0.35,0.5) {$\tilde{u}[k]$};
		% gain to loudspeaker beamformer
		\draw[->] (0,0) |- (0.9,1.25);
		%%%% microphones %%%%%%
		% microphone
		\draw (3.15,-2.25) circle (0.1);
		\draw (3.25,-2.1) -- (3.25,-2.4);
		%  microphone to gain
		\draw (3.05,-2.25) -| (0.0,-1.0);
		%%%% transfer path %%%%
		% draw transfer matrix of source to ear drum
		\draw (5.45,0) rectangle (6.55,-1.0);
		\node at (6,-0.5) {$H_{occ}(q)$};
		% draw connections from and to matrix
		\draw[->] (6.0,0) |- (6.35,1.15);
		% draw transfer matrix of ear canal
		\draw (4.0,0.75) rectangle (5.0,1.75);
		\node at (4.5,1.25) {${D}(q)$};
		% draw connections
		\draw[->] (3.5,1.25) -- (4.00,1.25); % 1st loudspeaker
		\draw[->] (5.0,1.25) -- (6.35,1.25);
		% draw transfer matrix of source to mic canal
		\draw (4.0,-2.75) rectangle (5.0,-1.75);
		\node at (4.5,-2.25) {$H_m(q)$};
		%%%% incoming signals
		\draw[->] (4.0,-2.25) -- (3.5,-2.25);
		% labels
		\node at (2.6,-2.05) {$y[k]$};
		\node at (2.6,0.95) {$u[k]$};
	\end{tikzpicture}	
 		\end{scriptsize}
	\caption{Generic single-microphone single-loudspeaker hearing device setup considered in this work.}
	\label{fig:simoframework}
\end{figure}
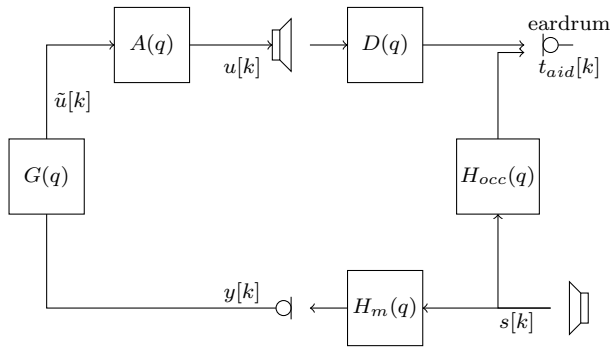

Consider a single-microphone single-loudspeaker hearing device depicted in Figure \ref{fig:simoframework}. We assume that all transfer functions are linear and time-invariant and can be modelled as polynomials in the variable $q$. Acoustic transparency is achieved when the aided transfer function of the individual ear $H_{aid}(q)$, consisting of the hearing device output and the sound leaking into the occluded ear, is (perceptually) equivalent to the open ear transfer function of the same ear where the desired hearing device processing is applied to, i.e.,
\begin{align}
	\underbrace{H_m(q)G(q)A(q)D(q) + H_{occ}(q)}_{H_{aid}(q)} = \underbrace{H_{open}(q)G(q)}_{H_{des}(q)}, \label{eq:transparency1}
\end{align}
where $H_{m}(q)$ is the transfer function from the source to the microphone of the hearing device, $G(q)$ is the desired hearing device processing, $A(q)$ is an equalization filter aiming to achieve acoustic transparency, $D(q)$ is the transfer function between the hearing device loudspeaker and the eardrum, $H_{occ}(q)$ is the transfer function between the source and the occluded eardrum and $H_{open}(q)$ is the transfer function between the source and the open eardrum. While the transfer function $G(q)$ of the hearing device processing can be assumed to be known, the transfer functions $H_{m}(q)$, $D(q)$, $H_{occ}(q)$, and $H_{open}(q)$ need to be measured or estimated individually. However, measurements of the transfer function to the individual eardrum are difficult to obtain, e.g., using probe-tube measurements. We can equivalently formulate \eqref{eq:transparency1} as
\begin{align}
	G(q)A(q)D(q) + \underbrace{\frac{H_{occ}(q)}{H_{m}(q)}}_{R_{occ}(q)} = \underbrace{\frac{H_{open}(q)}{H_{m}(q)}}_{R_{open}(q)}G(q). \label{eq:transparency2}
\end{align}
The optimal equalization filter in \eqref{eq:transparency2} depends on the following ratios:
\begin{enumerate}
	\item The ratio $R_{occ}(q)$ of the transfer function between the source and the occluded eardrum and the transfer function between the source and the hearing device microphone.
	\item The ratio $R_{open}(q)$ of the transfer function between the source and the open ear eardrum and the transfer function between the source and the hearing device microphone.
\end{enumerate}
Assuming knowledge of $G(q)$, the formulation in \eqref{eq:transparency2} allows to use estimated transfer functions $\hat{R}_{occ}(q)$ and $\hat{R}_{open}(q)$ and an individually estimated transfer function $\hat{D}(q)$ to compute the individual equalization filter $A(q)$. The main objective of this paper is to investigate the impact of different estimates $\hat{R}_{occ}(q)$, $\hat{R}_{open}(q)$, and $\hat{D}(q)$, on the equalized transfer function to the eardrum. In particular we investigate the impact of using an electro-acoustic model \cite{Vogl2019} to estimate the transfer function $D(q)$ from a measurement of the transfer function between the hearing device receiver and an in-ear microphone.

\section{Sound pressure equalization filter \label{sec:spleq}}
In this section, we briefly describe the sound pressure equalization filter computation. For more details, the interested reader is referred to \cite{Schepker2020jasm}. In order to compute the equalization filter $A(q)$ in \eqref{eq:transparency2} we use a least-squares based-design, where we aim at minimizing the following regularized least-squares cost function
\begin{align}
	J_{eq}(\mathbf{a}) = \Vert \mathbf{D}\mathbf{a} - (\mathbf{\hat{r}}_{open}-\mathbf{G}^{\dag}\mathbf{\hat{r}}_{occ})\Vert_2^2 + \lambda\Vert \mathbf{W} \mathbf{a} \Vert,
\end{align}
where $\mathbf{D}$ is the convolution matrix of $D(q)$ of appropriate dimensions and $\mathbf{a}$ is the $L_a$-dimensional vector of the finite impulse response equalization filter coefficients. The vectors $\mathbf{\hat{r}}_{open}$ and $\mathbf{\hat{r}}_{occ}$ are the finite impulse response approximation filter coefficients of $\hat{R}_{open}(q)$ and $\hat{R}_{occ}(q)$, $\mathbf{G}$ denotes the convolution matrix of the hearing device processing and $(\cdot)^\dag$ denotes the pseudo-inverse. The parameter $\lambda$ allows to trade-off between regularization and equalization and $\mathbf{W}$ is a weighting matrix to reduce the effect of comb-filtering \cite{Schepker2020jasm}.
The filter coefficients solution minimizing the cost function is computed as
\begin{align}
	\mathbf{a} = (\mathbf{D}^T\mathbf{D} + \lambda \mathbf{W}^T\mathbf{W})^{-1}\mathbf{D}^T(\mathbf{\hat{r}}_{open} - \mathbf{G}^\dag \mathbf{\hat{r}}_{occ}).
\end{align}
As mentioned above, one of the main objectives of this paper is to investigate the impact of different estimates $D(q)$ as well as $\mathbf{\hat{r}}_{occ}$ and $\mathbf{\hat{r}}_{open}$. For $D(q)$ we consider three different ways to compute these estimates. 
\begin{enumerate}
	\item Using individual estimates obtained using probe-tube measurements. While this is expected to yield the best equalization, it is time-consuming and hence often impractical.
	\item Using naive estimate obtained from an in-ear microphone. While is practically feasible, it is expected that this will results in suboptimal equalization since this transfer function is typically different from the transfer function to the eardrum
	\item Using an electro-acoustic model to estimate the tranfer function to the eardrum from an in-ear microphone. This is practically feasible and is expected to yield an equalization performance that is close to optimal.
\end{enumerate}
	Furthermore, we investigate two different ways of computing the estimates of $\mathbf{\hat{r}}_{occ}$ and $\mathbf{\hat{r}}_{open}$, either assuming availability of the individual measurements or the availability of multiple measurements from a database, i.e., 
\begin{enumerate}
	\item Using the individual estimate $\mathbf{\hat{r}}_{occ,ind}$ and $\mathbf{\hat{r}}_{open,ind}$, i.e.,
	\begin{align}
		\mathbf{\hat{r}}_{occ,ind} = \mathbf{H}_{m,ind}^\dag \mathbf{\tilde{h}}_{occ,ind}, \\
		\mathbf{\hat{r}}_{open,ind} = \mathbf{H}_{m,ind}^\dag \mathbf{\tilde{h}}_{open,ind},
	\end{align}
	where $\mathbf{H}_{m,ind}$ denotes the convolution matrix of $H_m(q)$ for the individual ear and $\mathbf{\tilde{h}}_{occ,ind}$ and $\mathbf{\tilde{h}}_{open,ind}$ denote the impulse response vector of $H_{occ}(q)$ and $H_{open}(q)$, respectively, zero-padded with $L_d$ leading zeros to allow for potential acausalities in the filter design. While this requires time-consuming and hence often impracticle measurement of $H_m(q)$, $H_{open}(q)$, and $H_{occ}(q)$, it is expected to provide the best possible equalization.
	\item Assuming the availability of $I$ a-priori measurements, e.g., for different subjects, of the transfer function between the source and the hearing device $H_{m,i}(q)$, $i=1,\dots,I$ ,the open ear transfer function $H_{open,i}(q)$, and the occluded ear transfer function  $H_{occ,i}(q)$, an average estimate of $\mathbf{\hat{r}}_{occ,av}$ and $\mathbf{\hat{r}}_{open,av}$ can be used, i.e.,
	\begin{align}
		\mathbf{\hat{r}}_{occ,av} = \Big(\sum_{i=1}^{I}\mathbf{H}_{m,i}^T\mathbf{H}_{m,i}^T\Big)^{-1} \sum_{i=1}^I \mathbf{H}_{m,i}^T\mathbf{\tilde{h}}_{occ,i}, \\
		\mathbf{\hat{r}}_{open,av} = \Big(\sum_{i=1}^{I}\mathbf{H}_{m,i}^T\mathbf{H}_{m,i}^T\Big)^{-1} \sum_{i=1}^I \mathbf{H}_{m,i}^T\mathbf{\tilde{h}}_{open,i},
	\end{align}
	where $\mathbf{H}_{m,i}$, is the convolution matrix of the $i$th measurement of $H_{m,i}(q)$, and $\mathbf{\tilde{h}}_{occ,i}$ and $\mathbf{\tilde{h}}_{open,i}$ are the impulse response vectors of the $i$th measurements of $H_{occ,i}(q)$ and $H_{open,i}(q)$, respectively, zero-padded with $L_d$ leading zeros. This practically feasible average estimate is expected to be more robust to variations in $H_{occ}(q)$ and $H_{open}(q)$ at the cost of a degraded accuracy.
\end{enumerate}

\section{Experimental Evaluation and Discussion}
In this section we experimentally evaluate the performance of the sound pressure equalization in terms of the aided magnitude responses at the eardrum, e.g., evaluating \eqref{eq:transparency1}, as well as objective sound quality predictions. Measurements of all acoustic transfer functions were conducted using a three-microphone two-receiver in-the-ear custom earpiece in an anechoic chamber for 12 different listeners (cf. also \cite{Vogl2019} for more on the data collection). While three different directions for the (external) source have been considered in \cite{Vogl2019}, here only the frontal direction is considered. Objective sound quality predictions were computed using the generalized power spectrum model quality (GPSMq) \cite{Biberger2018} which has shown to be highly correlated with subjective sound quality ratings of hear-through features for acoustic transparency \cite{Biberger2021}. For the GPSMq model we used the individual desired open-ear signal $G(q)H_{open}(q)s[k]$ as a reference signal and the aided ear signals $H_{aid}(q)s[k]$ as test signals. A total of eight different signals were used, 4 speech signals (2 male and 2 female) and 4 music signals (classical music, flute, jazz, piano).

In the first experiment, we investigate different estimates of the transfer function between the hearing device receiver and the eardrum $D(q)$. In the second experiment, we additionally investigate the impact of estimates of $R_{occ}(q)$ and $R_{open}(q)$, which results in practically feasible individualization. In all experiments we compute the equalization filter $A(q)$ as described in Section \ref{sec:spleq} using a filter length of $L_A=99$, a regularization constant of $\lambda = 10^{-1}$, an acausal delay of $L_d$ 32 samples, and a sampling rate of 16\,kHz. Furthermore, we use a hearing device gain of $G(q) = q^{-d_G}$, with $d_G$ a delay in samples, and the broadband amplification corresponding to a 0\,dB gain, i.e. a hearing device adjusted to achieve acoustic transparency. We investigate 4 different delays $d_G = \{0, 1, 16, 96\}$ samples, ranging from 0\,ms to 6\,ms.

\subsection{Experiment 1}
\begin{figure}
\figline{\leftfig{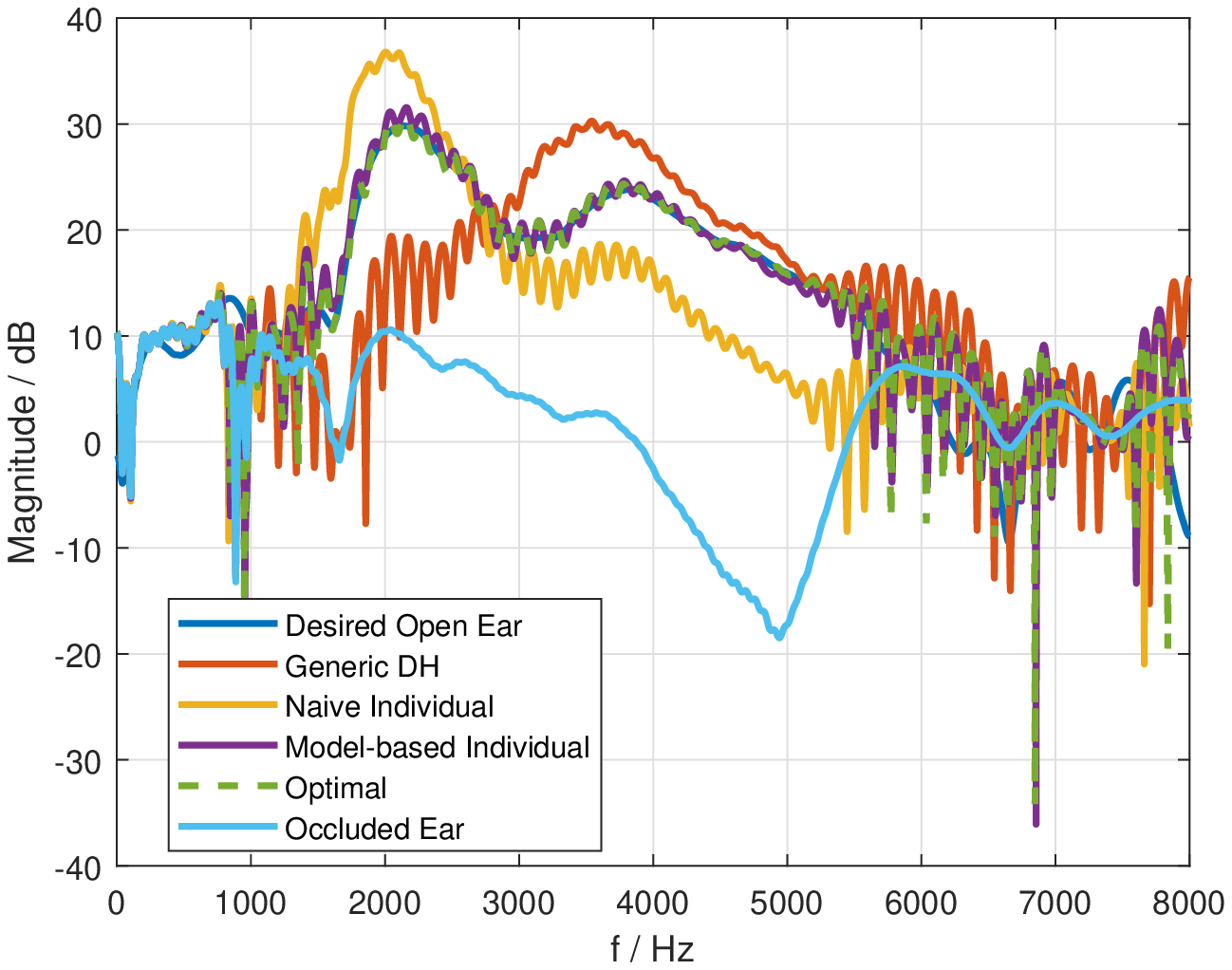}{.49\textwidth}{(a)} \leftfig{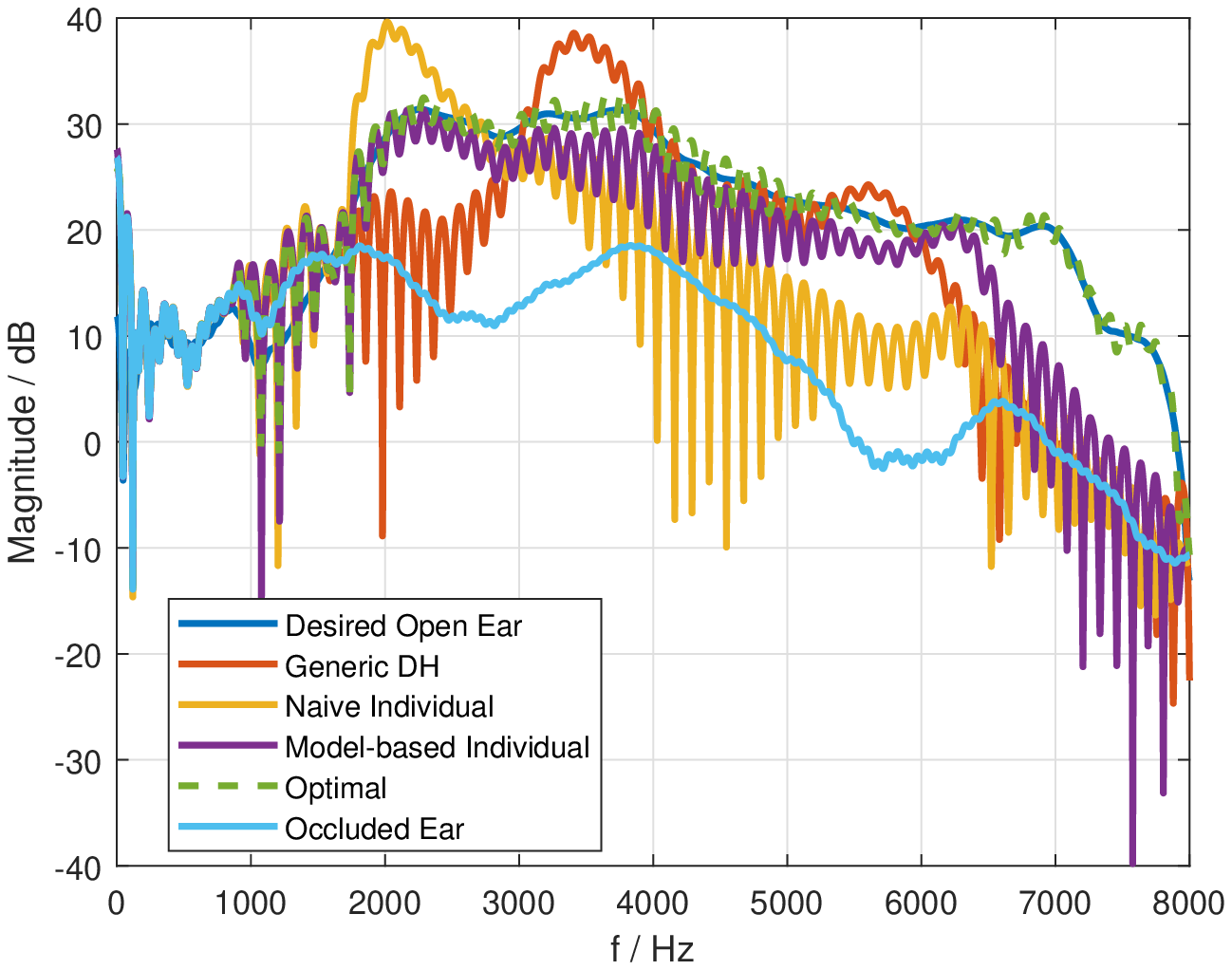}{.49\textwidth}{(b)}}
	\caption{Magnitude responses for experiment 1 using a hearing delay of 96 samples for the (a) best performing subject and (b) worst performing subject as measured by GPSMq predictions.}
	\label{fig:exp1_magnituderesp}
\end{figure}

In the first experiment, we investigate the impact of estimating the transfer function between the hearing device loudspeaker and the eardrum $D(q)$ on the equalization performance, and assume that all remaining transfer functions $H_m(q)$, $H_{open}(q)$, and $H_{occ}(q)$ are known (by measurement). To this end, we consider the following estimates:
\begin{itemize}
	\item {\it Optimal estimate:} we assume perfect knowledge of the individual transfer function (by probe-tube measurement).
	\item {\it Generic estimate DH:} we use measurements of all transfer functions performed on a dummy head to compute the equalization filter
	\item {\it Naive individual filter:} we use the individual transfer function between the receiver and the in-ear microphone.
	\item {\it Model-based individual filter:} we use the individualized electro-acoustic model \cite{Vogl2019} to estimate the transfer function between the receiver and the individual eardrum $D(q)$.
\end{itemize}
Figure \ref{fig:exp1_magnituderesp} shows magnitude responses for the desired open ear transfer function $G(q)H_{open}(q)$, the aided ear transfer function $H_{aid}(q)$ using the different estimates of $D(q)$, and the occluded ear transfer function $H_{occ}(q)$, using a delay of $d_G=96$. The subfigures show the magnitude responses for those subjects for which the best (Figure \ref{fig:exp1_magnituderesp}(a)) and worst (Figure \ref{fig:exp1_magnituderesp}(b)) sound quality is obtained as predicted by the GPSMq metric. Considering the best performing subject in Figure \ref{fig:exp1_magnituderesp}(a), it can be observed that the aided transfer functions differ substantially for the different estimates. Furthermore, all aided ear magnitude responses exhibit visible comb-filtering effects which are caused by the hearing device delay of 96\,ms. The optimal estimate using perfect knowledge of all transfer functions achieves a very close match to the desired open ear transfer function. Furthermore, the generic dummy head estimate exhibits large deviations from the desired open ear transfer function, in particular in the frequency region between 2kHz and 4kHz. This can be explained by the fact that the ear canal of the dummy head and the ear canal of the subject differ substantially. Similarly, using the Naive individual estimate yields differences compared to the desired open ear transfer function, in particular the resonances are overestimated (around 2kHz) and underestimated (around 4kHz). Using the model-based individual filter yields a very good approximation to the desired open ear transfer function without visible differences to the optimal estimate. Similar observations can be made for the Figure \ref{fig:exp1_magnituderesp}(b). Here it should be noted that the differences between the optimal estimate and the model-based individual estimate are considerably larger. This can be explained by the fact that for this subject the estimation of $D(q)$ is less accurate compared to the subject presented in Figure \ref{fig:exp1_magnituderesp}(a). 

\begin{figure}
\centering 
\includegraphics[width=0.49\textwidth]{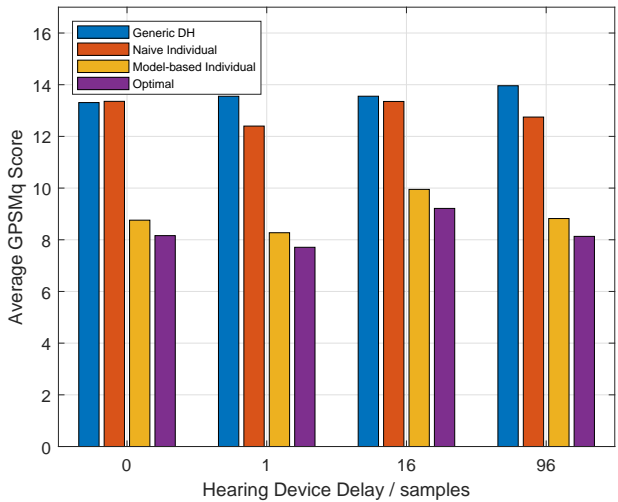}
	\caption{Sound quality predictions using the GPSMq measure for experiment 1.}
	\label{fig:exp1_quality}
\end{figure}

In order to investigate the impact of these equalization filters on the perceived quality, Figure \ref{fig:exp1_quality} shows sound quality predictions using the GPSMq metric averaged across all twelve subjects and eight audio signals. In general, the optimal estimate yields the best performance (i.e., the lowest GPSMq score). Furthermore, the model-based individual estimate yields only a small difference in the average GPSMq score compared to the optimal filter. In contrast, the generic DH filter and naive individual filter show a large degradation in sound quality compared to the optimal filter. Based on the magnitude responses shown in Figure \ref{fig:exp1_magnituderesp} this can be intuitively explained by the large differences in the magnitude responses between the aided ear transfer function and the desired open ear transfer function. These results demonstrate that a close-to-optimal individualized equalization can be achieved by using a model-based individual filter.

\subsection{Experiment 2}
\begin{figure}
\figline{\leftfig{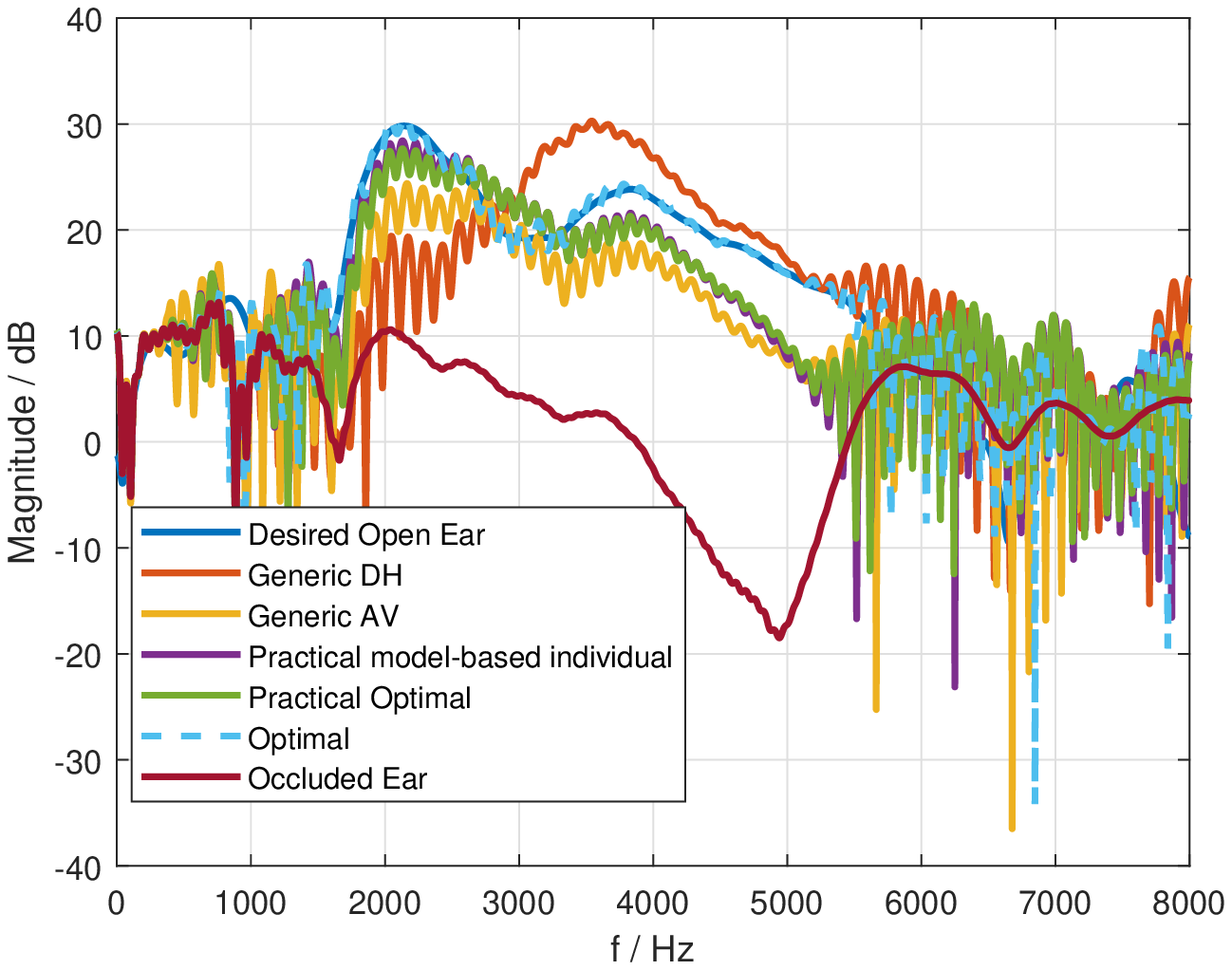}{.49\textwidth}{(a)} \leftfig{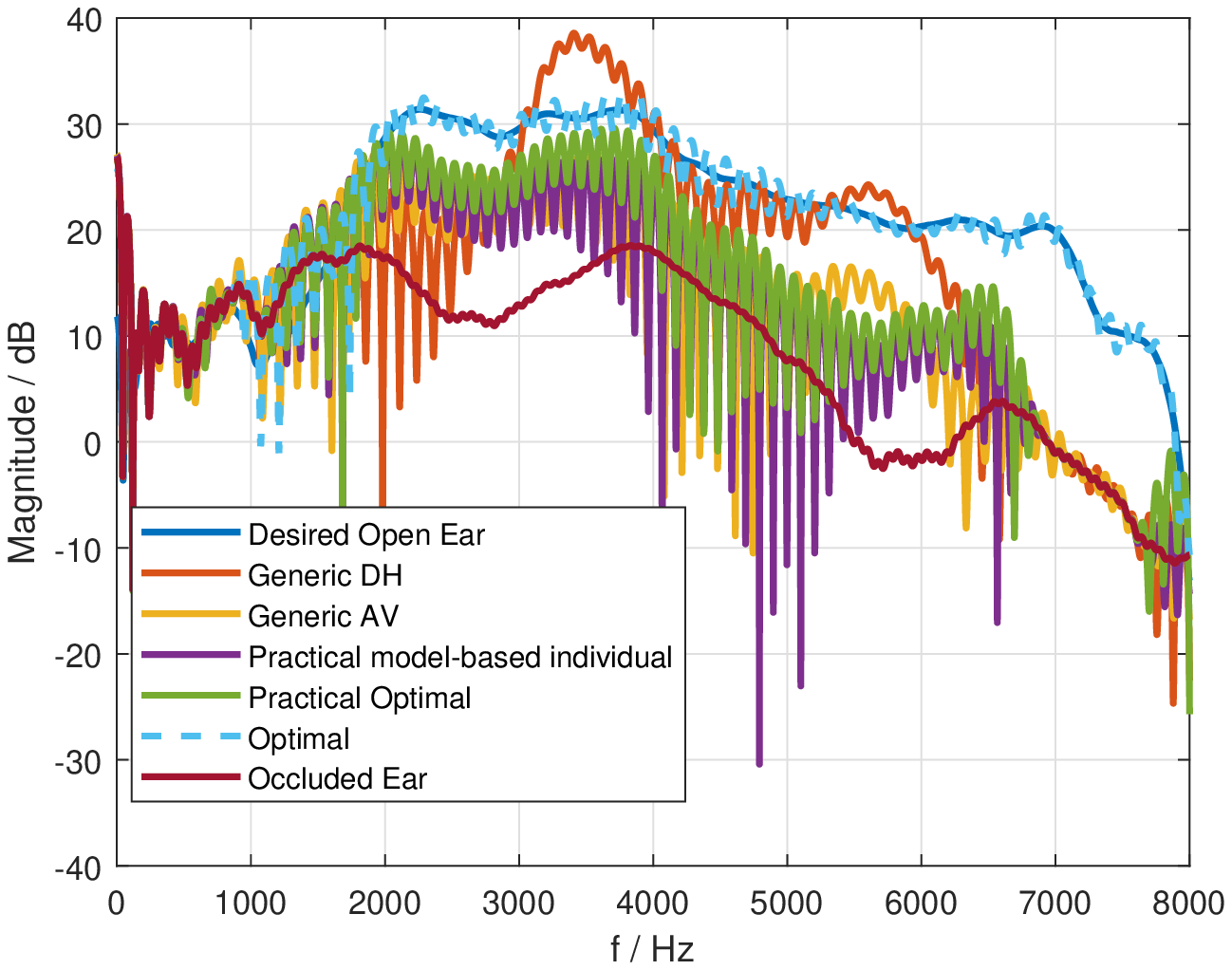}{.49\textwidth}{(b)}}
	\caption{Exemplary magnitude responses for experiment 2 using a hearing delay of 96 samples for the (a) best performing subject and (b) worst performing subject as measured by GPSMq predictions.}
	\label{fig:exp2_magnituderesp}
\end{figure}

\begin{figure}
\centering 
\includegraphics[scale=1.0]{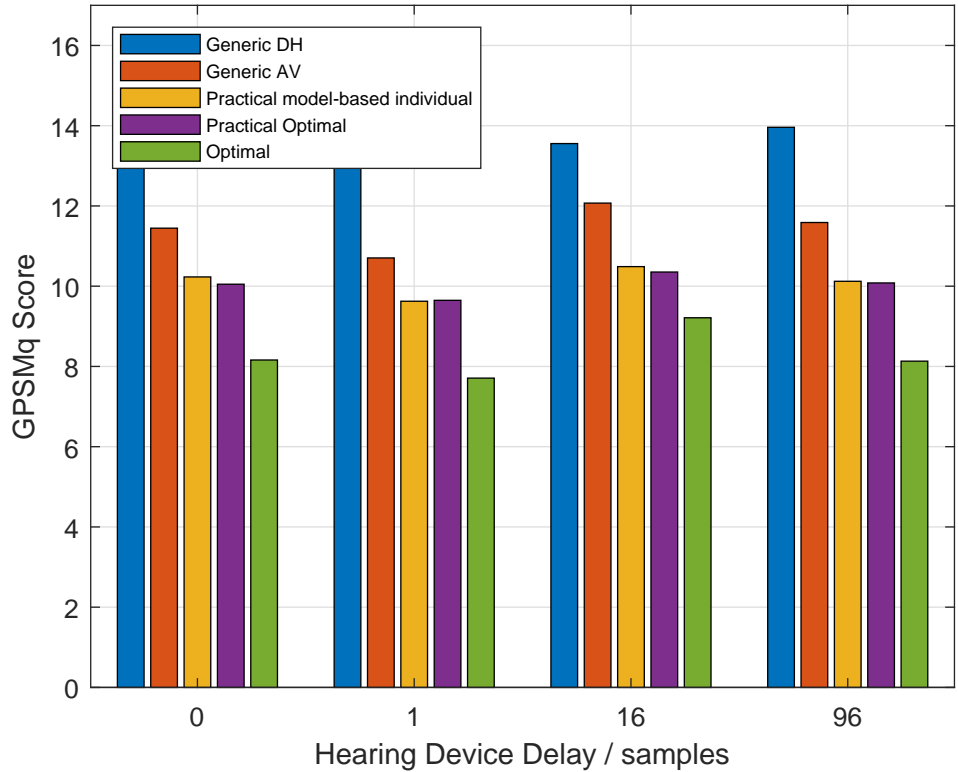}
	\caption{Sound quality predictions using the GPSMq measure for experiment 2.}
	\label{fig:exp2_quality}
\end{figure}

In the second experiment we investigate the impact of estimating the transfer functions between the source and the hearing device and between the source and the open and occluded eardrum. We assume that the transfer function between the hearing device loudspeaker and the eardrum is either known by measurement or is estimated from an in-ear microphone using the individualized electro-acoustic model. Specifically, the following estimates are considered:
\begin{itemize}
	\item {\it Optimal:} we assume perfect knowledge of all individual transfer functions.
	\item {\it Generic filter DH:} we use measurements performed on a dummy head to compute the equalization filter.
	\item {\it Generic filter AV:} we use multiple measurements to obtain a filter that is optimal on average across a set of individual measurements (the subject under test was not included in this set of measurements).
	\item {\it Practical model-based individual:} we use multiple measurements to compute average estimates of $R_{open}(q)$ and $R_{occ}(q)$ (the subject under test was not included in this set of measurements) and use $\hat{D}(q)$ estimated using the individualized electro-acoustic model.
	\item {\it Practical optimal:} we use multiple measurements to compute average estimates of $R_{open}(q)$ and $R_{occ}(q)$ (the subject under test was not included in this set of measurements) and the measured individual estimate of $D(q)$.
\end{itemize}
Figure \ref{fig:exp2_magnituderesp} shows magnitude responses for the desired open ear transfer function $G(q)H_{open}(q)$, the aided ear transfer function $H_{aid}(q)$ using the different estimates of $D(q)$, and the occluded ear transfer function $H_{occ}(q)$, using a delay of $d_G=96$. The subfigures show the magnitude responses for those subjects for which the best (Figure \ref{fig:exp2_magnituderesp}(a)) and worst (Figure \ref{fig:exp2_magnituderesp}(b)) sound quality is obtained as predicted by the GPSMq metric. Similarly as in Experiment 1, considering the best performing subject in Figure \ref{fig:exp2_magnituderesp}(a), it can be observed that the aided transfer functions differ substantially for the different estimates and comb-filtering effects are visible. It should be noted that the optimal filter and the generic dummy head filter have been already discussed in Experiment 1. Using a generic average filter yields a better match to the desired open ear transfer function compared to the generic DH filter. In contrast, as can be observed for both the practical model-based individual and the practical optimal filter, including individual ear information yields an even closer match to the desired open ear transfer function. Similarly as in Experiment 1, the differences between magnitude responses of the optimal estimate and the model-based individual estimate are considerably smaller compared to the Naive individual estimate. Similar observations are made for Figure \ref{fig:exp2_magnituderesp}, though generally the difference between the desired open ear and the aided cases is larger. Again the practical model-based individual filter shows the closest match to practical optimal filter. It should be noted that for this subject also the generic average filter yields similar equalization performance.
In order to investigate the impact of these equalization filter on the perceived quality, Figure \ref{fig:exp2_quality} shows average sound quality predictions using the GPSMq metric. In general, the optimal estimate yields the best performance (i.e., the lowest GPSMq score). Furthermore, the practical optimal and practical model-based individual estimate yield similar performance and achieve the best performance amongst the practical filters. In contrast, the generic DH and generic AV estatimes lead to a large degradation in sound quality compared to the optimal filter. This can be explained by the large differences in the magnitude responses between the aided ear transfer function and the desired open ear transfer function as observed in Figure \ref{fig:exp2_magnituderesp}. These results demonstrate that practical individualized sound pressure equalization can be achieved by using the practical model-based individual filter.

\section{Conclusion}
In this paper different approaches for individualized sound pressure equalization were experimentally compared by means of aided transfer functions and sound quality predictions. In particular individualization using an electro-acoustic model of the hearing device was carried out to estimate the sound pressure at the eardrum. Incorporating these individual estimates into the sound pressure qualization filter computation allowed to achieve significant improvements in sound quality compared to generic estimates as predicted by an objective sound quality metric. Furthermore, using average estimates for the open ear transfer function and the occluded ear transfer function together with the individual estimate of the transfer function between the hearing device and the eardrum yielded practical sound pressure equalization filters achieving better predicted sound quality compared to using a generic average filter. These results demonstrate that considering individual ear acoustics with particular focus on the sound pressure at the individual eardrum is important when equalizing the sound pressure in hearing devices. Further studies are required to substantiate these findings with perceptual data, e.g., using sound quality judgements.
% If only one acknowledgment:
%\begin{acknowledgment}
%This research was supported by  ...
%\end{acknowledgment}

%or

\begin{acknowledgments}
This work was funded by the Deutsche Forschungsgemeinschaft (DFG, German Research Foundation) – Project ID 352015383 (SFB 1330 A4 and C1), and Project ID 390895286 (EXC 2177/1).
\end{acknowledgments}

% -------------------------------------------------------------------------------------------------------------------
%  No Appendix in JASA-EL

%=======================================================

% Use \bibliography{<name of your .bib file>}+

\begin{thebibliography}{1}%
\bibitem[Biberger et al.(2018)]{Biberger2018}T.~Biberger, J.-H.~Fle{\ss}ner, R.~Huber, S.~D.~Ewert, "An objective audio quality measure based on power and envelope power cues," \emph{J. Audio Eng. Soc.}, vol. 66, no. 7/8, pp. 578--593, July 2018.
%
\bibitem[Biberger et al.(submitted)]{Biberger2021}T.~Biberger, H.~Schepker, F.~Denk, S.~Ewert, "Instrumental quality predicitions and analysis of auditory cues for algorithms in modern headphone technology," \emph{Trends in Hearing}, in revision.
%
\bibitem[Denk et al.(2018a)]{Denk2018}F. Denk, M. Hiipakka, B. Kollmeier, and S. M. A. Ernst, "An individualised acoustically transparent earpiece for hearing devices," \emph{Int. J. Aud.}, vol. 57, no. suppl. 3, pp. 62--70, 2018.

\bibitem[Denk et al.(2018b)]{Denk2018TIH}F. Denk, S. M. A. Ernst, and S. D. Ewert, and B. Kollmeier, "Adapting Hearing Devices to the Individual Ear Acoustics: Database and Target Response Correction Functions for Various Device Styles," \emph{Trends in Hearing}, vol. 22, pp. 1--19, 2018.
%
\bibitem[Fabry et al.(2019)]{Fabry2019}J.~Fabry, F.~K{\"o}nig, S.~Liebich, P.~Jax, "Acoustic equalization for headphones using a fixed feed-forward filter," in \emph{Proc. IEEE Int. Conf. Acoustics, Speech, Signal Process. (ICASSP)}, Brighton, United Kingdom, May 2019, pp. 980-984.
%
\bibitem[Gupta et al.(2019)]{Gupta2019}R. Gupta, R. Ranjan, J. He, G. Woon-Seng, "Parametric hear through equalization for augmented reality audio," Proc. Int. Conf. Acoustc., Speech, Signal Process. (ICASSP), Brighton, United Kingdom, May 2019, pp. 1587--1591.
%
\bibitem[Hellstrom and Axelsson(1993)]{Hellstrom1993}P.-A.~Hellstrom and A.~Axelsson, "Miniature microphone probe tube measurements in the external auditory canal," \emph{J. Acoust. Soc. Am.}, vol. 93, no. 2, pp. 907--919, Feb. 1993.
%
\bibitem[Hoffmann et al.(2013)]{Hoffmann2013}P.~Hoffmann, F.~Christensen, and D.~Hammersh{\o}i, "Insert earphone calibration for hear-through options," in Proc. AES Conf.: Loudspeaker \& Headphones, Aalborg, Denmark, Aug. 2013, pp. 3--4.
%
\bibitem[Hudde et al.(1999)]{Hudde1999}H.~Hudde, A.~Engle, and A.~Lodwig, "Methods for estimating the sound pressure at the eardrum," \emph{J. Acoust. Soc. Am.}, vol. 106, no. 4 (Pt. 1), 1977--1992, Oct. 1999.
%
\bibitem[Killion(2004)]{Killion2004}M. C. Killion, "Myths that discourage improvements in hearing aid design," \emph{Hearing Review}, vol. 11, no. 4, pp. 32--70, Jan. 2004.
%
\bibitem[McCreery et al.(2009)]{McCreery2009}R.~W.~McCreery, A.~Pittman, J.~Lewis, S.~T.~Neely, P.~G.~Stelmachowicz, "Use of forward pressure level to minimize the influence of acoustic standing waves during probe-microphone hearing-aid verification," emph{J. Acoust. Soc. Am.}, vol. 126, no. 1, pp. 15--24, Jul. 2009.
%
\bibitem[Sankowsky-Rothe et al.(2015)]{SankowskyRothe2015}T.~Sankowsky-Rothe, M.~Blau, S.~K\"ohler, A.~Stirnemann, "Individual equalization of hearing aids with integrated ear canal microphones," \emph{Acta Acustica united with Acustica}, vol. 101, no. 3, pp. 552--566, 2015.
%
\bibitem[Schepker et al. (2020)]{Schepker2020jaes}H.~Schepker, F.~Denk, B.~Kollmeier, S.~Doclo, "Acoustic transparency in hearables -- perceptual sound quality evaluations," \emph{J. Audio Eng. Soc.}, vol. 68, no. 7/8, pp. 495--507, Jul./Aug. 2020.
%
\bibitem[Schepker et al.(submitted)]{Schepker2020jasm}H.~Schepker, F.~Denk, B.~Kollmeier, S.~Doclo, "Robust single- and multi-loudspeaker least-squares-based equalization for hearing devices," \emph{arXiv:2109.04241 [eess.AS]}, Sep. 2021.
%
\bibitem[Sockalingam et al.(2009)]{Sockalingam2009}R. Sockalingam, J. Beilin, and D. L. Beck, "Sound quality considerations of hearing instruments," \emph{Hearing Review}, vol. 16, no. 9, pp. 22--28, 2009.
%
\bibitem[Stinson(1985)]{Stinson1985}M.~R.~Stinson, "The spatial distribution of sound pressure within scaled replicas of the human ear canal," \emph{J. Acousti. Soc. Am.}, vol 78, no. 5, pp. 1596--1602, Nov. 1985.
%
\bibitem[V{\"a}lim{\"a}ki et al.(2015)]{Vaelemaeki2015}V. V{\"a}lim{\"a}ki, A. Franck, J. R{\"a}m{\"o}, H. Gamper, L. Savioja, "Assisted Listening using a Headset," \emph{Signal Process. Mag.}, vol. 32, no. 2, Mar. 2015, pp. 92--99.
%
\bibitem[Vogl \& Blau(2019)]{Vogl2019} S. Vogl, and M. Blau, "Individualized prediction of the sound pressure at the eardrum for an earpiece with integrated receivers and microphones," \emph{J. Acoust. Soc. Am.}, vol. 145, no. 2, pp. 917--930, Feb. 2019.
\end{thebibliography}
% to make your bibliography with BibTeX. 

% Run BibTeX using the name of your .tex file. This will produce
% a filename.bbl file.

% Then run pdfLaTeX several times on the .tex file, to get the
% bibliography and hyperlinked citations.

%=======================================================
\
\end{document}